# Comments on "Mixed Bose-Fermi statistics Kinetic equation and navigation through network" by S.F. Chekmarev, Phys. Rev. E 82, 026106 (2010)

I.H. Umirzakov
*Institute of Thermophysics, Novosibirsk, Russia*

**Abstract**

The paper shows that the kinetic equations considered in [1], equilibrium distribution obtained in [1], and results and conclusions obtained on the basis of the kinetic equation derived in [1] do not correspond to the mixed Bose-Fermi statistics. Moreover, it is shown that the kinetic equation corresponding to the case when the copies of the system are characterized by different values of the fraction of the Fermi-like moves is incorrect.

We present a correct kinetic equation for the mixture of the Bose and Fermi moves and obtained the equilibrium distribution for the case when the probability of the Fermi moves is higher or equal to that of the Bose moves.

**Introduction**

The conventional master equation was generalized to include the Bose and Fermi moves in [1]. According to [1], the obtained equation describes the time evolution of a system with the mixed Bose-Fermi statistics which is controlled by the fraction of Fermi moves, the latter plays a role of an exploratory tendency (ET) factor. The procedure used to include the Bose and Fermi moves in the master equation offers a framework for derivation of kinetic equations for more complex and general statistics. The theoretical consideration was illustrated with numerical results for navigation through a model scale-free network [1]. Depending on the ET factor and the number of participants involved in the navigation a broad variety of the behavior scenarios has been observed. The cases of one participant and many participants have been found drastically different.

These Comments show that the kinetic equations considered in [1], equilibrium distribution obtained in [1], and results and conclusions obtained on the basis of the kinetic equation derived in [1] do not correspond to the mixed Bose-Fermi statistics. Hence, the title "Mixed Bose-Fermi statistics: Kinetic equation and navigation through a network" does not correspond to the content of [1]. We also show that the kinetic equation corresponding to the case when the copies of the system are characterized by different values of the fraction of the Fermi-like moves is incorrect. Moreover, we derive the correct kinetic equations for a mix of the Bose and Fermi moves and obtain the equilibrium distribution for the case when the probability of the Fermi moves is higher or equal to that of Bose moves.

## "Mixed Bose-Fermi statistics" [1]

To start with, we present the main equations (1)-(17) which were considered in [1].
The kinetic equation

$$\frac{dN_i}{dt} = \sum_j [k_{i,j} N_j(t) - k_{j,i} N_i(t)], \tag{1}$$

was considered. Here, $N_i(t)$ is the mean number of copies of the system in state $i$, which obeys the condition

$$\sum_i N_i(t) = N_{tot}, \tag{2}$$

$i = 0, 1, 2, \ldots$, $N_{tot}$ is the total number of copies of the system, $k_{i,j}$ is the rate of transitions from state $j$ to $i$, obeying the detailed balance

$$k_{i,j} N_j^0 = k_{j,i} N_i^0, \tag{3}$$

where

$$N_i^0 = N_{tot} \exp(-\beta E_i)/Z, \tag{4}$$

$E_i$ is the energy of state $i$ (the states are assumed to be non-degenerate), $\beta = \frac{1}{k_B T}$ is the inverse temperature, $k_B$ is the Boltzmann constant, $T$ is the temperature, and $Z = \sum_i \exp(-\beta E_i)$ is the partition function. Using the equality

$$k_{i,j} N_j(t) = k_{i,j}[1 - N_i(t)] N_j(t)/2 + k_{i,j}[1 + N_i(t)] N_j(t)/2, \tag{5}$$

Eq. (1) was presented as

$$\frac{dN_i}{dt} = \frac{1}{2} \sum_j \{k_{i,j}[1 - N_i(t)] N_j(t) - k_{j,i}[1 - N_j(t)] N_i(t)\}$$

$$+ \frac{1}{2} \sum_j \{k_{i,j}[1 + N_i(t)] N_j(t) - k_{j,i}[1 + N_j(t)] N_i(t)\}. \tag{6}$$

According to [1]:
- the first term in Eq. (5) can be associated with a Fermi-like move, and the second term with a the Bose-like move;
- the first sum in the right-hand side of Eq. (6) presents the Fermi-like moves and the second one presents the Bose-like moves;
- the evolution of the system that is described by the conventional master equation (1) "can be considered as a repeated sequence of two sufficiently small and, (strictly speaking, infinitesimal) half-steps: in one half-step (a Fermi-like move), the system samples the state space, favoring the states that have not been visited yet, and in the other (a Bose-like move), it returns to a previously visited state, favoring the states that have been occupied with higher probabilities (a reasonable strategy to avoid getting lost in an unknown place)";
- the equation

$$k_{i,j}[1 - N_i(t)] N_j(t) = k_{j,i}[1 - N_j(t)] N_i(t) \tag{7}$$

of detailed balance for the Fermi moves gives

$$N_i^F = \frac{1}{C_F \exp(\beta E_i) + 1}, \tag{8}$$

where $C_F$ is defined from the condition

$$\sum_i N_i^F = N_{tot}; \qquad (9)$$

- the equation

$$k_{i,j}[1 + N_i(t)]N_j(t) = k_{j,i}[1 + N_j(t)]N_i(t) \qquad (10)$$

of detailed balance for the Bose moves gives

$$N_i^B = \frac{1}{C_B \exp(\beta E_i) - 1}, \qquad (11)$$

where $C_B$ is defined from the condition

$$\sum_i N_i^B = N_{tot}; \qquad (12)$$

- Eq. (6) was generalized as

$$\frac{dN_i}{dt} = \alpha \sum_j \{k_{i,j} L_i [1 - N_i(t)] N_j(t) - k_{j,i} L_j [1 - N_j(t)] N_i(t)\}$$
$$+ (1 - \alpha) \sum_j \{k_{i,j}[1 + N_i(t)] N_j(t) - k_{j,i}[1 + N_j(t)] N_i(t)\}, \qquad (13)$$

where $\alpha$ is the fraction of the Fermi-like moves and $L_i = \theta(1 - N_i)$, where $\theta(x)$ is the Heaviside step function: $\theta(x) = 0$ for $x < 0$, and $\theta(x) = 1$ for $x \geq 0$, so $L_i = 1$ for $N_i \leq 1$ and $L_i = 0$ for $N_i > 1$;

- the equation

$$k_{i,j}\{\alpha L_i[1 - N_i(t)]N_j(t) + (1 - \alpha)[1 + N_i(t)]N_j(t)\}$$
$$= k_{j,i}\{\alpha L_j[1 - N_j(t)]N_i(t) + (1 - \alpha)[1 + N_j(t)]N_i(t)\}, \qquad (14)$$

of detailed balance corresponding to Eq. (13) gives

$$N_i^{BF} = \frac{1 + (L_i - 1)\alpha}{C_{BF} \exp(\beta E_i) + (L_i + 1)\alpha - 1}, \qquad (15)$$

where $C_{BF}$ is defined from the condition

$$\sum_i N_i^{BF} = N_{tot}; \qquad (16)$$

- Eq. (13) assumes all copies of the system to be identical, i.e. they have the same $\alpha$ and start at the same time and state; and

- in the case when the copies of the system are characterized by different values of $\alpha$ and/or start at different times and states, the kinetic equation (13) writes as

$$\frac{dN_i^{(m)}}{dt} = \alpha^{(m)} \sum_j \{k_{i,j} L_i [1 - N_i(t)] N_j^{(m)}(t) - k_{j,i} L_j [1 - N_j(t)] N_i^{(m)}(t)\}$$
$$+ (1 - \alpha^{(m)}) \sum_j \{k_{i,j}[1 + N_i(t)] N_j^{(m)}(t) - k_{j,i}[1 + N_j(t)] N_i^{(m)}(t)\}. \qquad (17)$$

## Comments

Below we present our comments to [1].

1. According to Eqs. (2), (9), (12) and (16), the sums $\sum_i$ mean the summation over all states. Therefore, the kinetic equations (1), (6), (13) and (17) are incorrect, because in the right-hand sides of them the sums $\sum_j$ are over all states, while the state $j = i$ must be excluded [2-5]. Therefore, the sums $\sum_j$ in the right-hand sides of Eqs. (1), (6), (13) and (17) must be replaced by $\sum_{j, j \neq i}$.

2. According to [1], Eq. (13) assumes that all copies of the system start at the same time and state. It is evident that Eq. (13) can correspond to the cases when all copies start at the same time from different initial states or at various times from the same state or at various times from different initial states.

3. Eqs. (7), (10) and (14) are incorrect because the mean numbers $N_i$ does not depend on time in equilibrium conditions. These incorrect equation must be replaced by the following correct ones:

$$k_{i,j}(1 - N_i^F)N_j^F = k_{j,i}(1 - N_j^F)N_i^F,$$

$$k_{i,j}(1 + N_i^B)N_j^B = k_{j,i}(1 + N_j^B)N_i^B,$$

$$k_{i,j}\left[\alpha(1 - N_i^{BF})N_j^{BF} + (1-\alpha)(1 + N_i^{BF})N_j^{BF}\right] = k_{j,i}\left[\alpha(1 - N_j^{BF})N_i^{BF} + (1-\alpha)(1 + N_j^{BF})N_i^{BF}\right],$$

respectively.

4. Eq. (17) is incorrect because $N_i(t)$, $N_j(t)$ and $N_j^{(m)}(t)$ in the right-hand side of it must be replaced by $N_i^{(m)}(t)$, $N_j^{(m')}(t)$ and $N_j^{(m')}(t)$, respectively, and the sums $\Sigma_j$ must be replaced by $\Sigma_{j,j\neq i}\Sigma_{m'}$.

5. According to [1], the case $\alpha = 1$ in Eqs. (13)-(15) corresponds to the Fermi statistics. We have from Eq. (15) for $\alpha = 1$

$$N_i^{BF} = \frac{L_i}{C_{BF}\exp(\beta E_i) + L_i}. \tag{18}$$

The comparison of Eqs. (8) and (18) shows that Eq. (18) does not correspond to the Fermi statistics. Hence, Eqs. (13)-(15) do not concern the Fermi statistics in this case.

6. According to [2-6], $N_{tot}$ is the number of the particles in the Bose and Fermi ideal gases. In the case when $N_{tot} = 1$ the equilibrium distribution of one boson or fermion is given by [3]

$$N_i^1 = \exp(-\beta E_i) / \Sigma_i \exp(-\beta E_i). \tag{19}$$

In this case we have $N_i(t) \leq 1$ and $L_i = 1$. Eq. (15) gives

$$N_i^{BF} = \frac{1}{C_{BF}\exp(\beta E_i) + 2\alpha - 1}. \tag{20}$$

The comparison of Eqs. (19) and (20) shows that Eq. (19) does not correspond to both the Bose ($\alpha = 0$) and Fermi ($\alpha = 1$) statistics in [1]. Hence, Eqs. (13)-(15) are incorrect in the case when one boson or fermion is under consideration.

Note that the equilibrium Fermi and Bose distributions given by Eqs. (8)-(9) and (11)-(12) correspond to $N_{tot} \gg 1$ [2-5].

7. According to [1], $N_i^{BF} \ll 1$ is true for $N_{tot} = 1$. However, $N_i \leq 1$ is valid for $N_{tot} = 1$. Therefore, $N_i^{BF} \ll 1$ is not valid for $N_{tot} = 1$ in the general case.

8. One can conclude from above considerations that if $N_{tot} = 1$, then Eqs. (13)-(15) are incorrect in the case when $\alpha \neq 1/2$.

9. We can conclude from above consideration of the case $N_{tot} = 1$ that the following statements in [1] such as:

1) "At the same time, if $\alpha = 1$ (solely Fermi moves, no possibility to return to a previously visited state), only one of the participants attains the target state but it does it much faster than when a single copy navigates through the network ($N_{tot} = 1$), i.e., this one optimizes the way to the target state at the expense of other participants"; 2) "For $N_{tot} = 1$, where $N_i^{BF} \ll 1$, the statistics are close to the Boltzmann statistics; in particular, the distribution at $\alpha = 0.5$ is the true Boltzmann distribution"; 3) "At $N_{tot} = 1$ the behavior of the system changes with $\alpha$ in a regular way, i.e., as $\alpha$ increases, the system explores a larger portion of the state space and, consequently, spends a longer time to reach the target state [Figs. 2(a), 3(a), and 4]"; 4) "It is of interest that in this case the MFPT is much shorter than the MFPT at $N_{tot} = 1$ ( ~10 times), Fig. 4)"; 5) "In the case of $N_{tot} = 1$ neither the separation of the Fermi and Bose fluxes nor the critical slowdown of the fluxes is observed"; 6) "At $\alpha = 1$, as previously for the ensemble of identical copies, the target state is occupied by a single copy, and the other copies assist it in reaching the target state faster than in the case of a single copy $N_{tot} = 1$"; and finally, 7) "A drastic difference between the cases of $N_{tot} = 1$ and $N_{tot} \gg 1$ has been observed" have no physical sense.

10. One can also conclude that the population of the target state with time for $N_{tot} = 1$ in Fig. 2 [1] and distributions of the first passage time to the target state for $N_{tot} = 1$ in Fig. 3 [1], as well as the mean first passage times versus the exploratory tendency factor $\alpha$ corresponding to $N_{tot} = 1$ in Fig. 4 [1] are incorrect.

11. The conclusions in [1] made on the basis of the comparison of the cases $N_{tot} = 1$ and $N_{tot} \gg 1$, specifically $N_{tot} = 10$, are incorrect.

12. The kinetic equation of the ideal Fermi gas consisting of $N_{tot}$ ($N_{tot} > 1$) fermions is

$$\frac{dN_i}{dt} = \sum_{j, j \neq i} \{k_{i,j} L_i [1 - N_i(t)] L_j N_j(t) - k_{j,i} L_j [1 - N_j(t)] L_i N_i(t)\}, \tag{21}$$

where the condition

$$N_i(t) \leq 1 \tag{22}$$

for fermions is taken into account [2-5].

According to [1], the first sums in the right-hand sides of Eqs. (6), (13) and (17) correspond to the Fermi-like moves. The comparison of these sums with the sum in the right-hand side of Eq. (21) shows that these sums in the right-hand sides of Eqs. (6), (13) and (17) do not correspond to the Fermi-like moves.

13. According to [1], the Fermi flux to the target state is determined as $J_F(t) = \alpha \sum_j k_{T,j} [1 - N_T(t)] L_T N_j(t)$. This definition of the Fermi flux is incorrect. According to a correct kinetic equation (21), the correct Fermi flux is defined as

$$J_F(t) = \alpha \sum_j k_{T,j} [1 - N_T(t)] L_T N_j(t) L_j.$$

14. One can also conclude that the data on the Fermi flux to the target state given on Fig. 5 [1] are incorrect. Therefore, the conclusions made in [1] on basis of these data are incorrect.

15. Using the comments above, one can conclude that the following statements in [1] such as:

1) "The conventional master equation is generalized to include Bose and Fermi moves"; 2) "The obtained equation describes the time evolution of a system with mixed Bose-Fermi statistics which are controlled by the fraction of Fermi moves"; 3) "The procedure used to include Bose and Fermi moves in the master equation offers a framework for derivation of kinetic equations for more complex and general statistics"; 4) "In this paper, the conventional master equation is generalized to include Bose and Fermi moves"; 5) "The obtained equation describes the time evolution of a system with mixed Bose-Fermi statistics"; 6) "the fraction of Fermi moves serves as an exploratory tendency ET factor"; 7) "In contrast with Refs. [12,13], the mixed statistics interpolate between Bose and Fermi statistics in an explicit way through the fraction of Fermi moves"; 8) "The procedure used to include Bose and Fermi moves offers a framework to derive kinetic equations for more complex and general statistics"; 9) "Let us split every $k_{m,l}N_l(t)$ term in Eq. (1) ($m,l = i,j$) in two parts as follows: $k_{m,l}N_l(t) = k_{m,l}[1 - N_m(t)]N_l(t)/2 + k_{m,l}[1 + N_m(t)]N_l(t)/2$. The first term in this equality, in which the probability of transitions to state $m$ is proportional to $1 - N_m(t)$, can be associated with a Fermi-like move"; 10) "After the corresponding rearrangement, Eq. (1) writes as

$$\frac{dN_i}{dt} = \frac{1}{2}\sum_j \{k_{i,j}[1 - N_i(t)]N_j(t) - k_{j,i}[1 - N_j(t)]N_i(t)\}$$

$$+ \frac{1}{2}\sum_j \{k_{i,j}[1 + N_i(t)]N_j(t) - k_{j,i}[1 + N_j(t)]N_i(t)\}, \qquad (4)$$

where the first sum in the right-hand side of the equation presents Fermi-like moves"; 11) "It follows that the evolution of the system that is described by the conventional master equation, i.e., Eq. (1), and leads to the Boltzmann distribution at equilibrium can be considered as a repeated sequence of two sufficiently small (strictly speaking, infinitesimal) half-steps: in one half-step (a Fermi-like move), the system samples the state space, favoring the states that have not been visited yet, and in the other (a Bose-like move), it returns to a previously visited state, favoring the states that have been occupied with higher probabilities (a reasonable strategy to avoid getting lost in an unknown place)"; 12) "$\alpha$ is the fraction of Fermi-like moves"; 13) "The step function is introduced to restrict further the growth of population of the states in Fermi-like moves when the occupation numbers $N_m(t)$ have reached unity, which makes it possible to account for congestion effects"; 14) "We also note that at $N_m(t) > 1$ the effective rates of transitions for Fermi-like moves, $k_{m,l}[1 - N_m(t)]$, would be negative"; 15) "At $N_{tot} \gg 1$, the distribution varies substantially from the Bose distribution at $\alpha = 0$ to the Fermi one at $\alpha = 1$"; 16) "As $\alpha$ becomes closer to unity, the distributions deviate from the exponential distributions considerably, particularly at $\alpha = 1$ (solely Fermi moves), when the system favors visiting any new state, i.e., it performs a sort of the breadth-first search"; 17) "At $\alpha = 1$ (solely Fermi moves), the system reaches the target state very fast, considerably faster than in all other cases (Fig. 4), but the population of the target state is limited by the value of $N_T = 1$ [Fig. 2(b)], i.e., only one copy attains the target state"; 18) "Because the state space is partly occupied by the other copies, an individual copy samples a less portion of the space (due to the Fermi effect) and thus optimizes the way to the target state"; 19) "At $\alpha = 1$ (mixed Bose and Fermi moves), the target state is completely populated with time ($N_T = N_{tot} = 10$), i.e., all copies attain the target state"; 20) "However, this may require a much longer time than for a single copy, because as $N_T$ becomes close to unity, its further occupation is difficult"; 21) "Figure 5 shows these fluxes at $\alpha = 0.5$ and $\alpha = 0.995$. In each case, as $N_T(t)$ reaches unity [cf. Fig. 2(b)], the Fermi flux terminates, and the further occupation of the target state is provided by the Bose flux"; 22) "Correspondingly, the FPT distribution, which can be written as $p_T(t) = J_F(t) + J_B(t)$, has two local maxima at the points where $J_F(t)$ and $J_B(t)$ have maximum values [Fig. 3(b)]"; 23) "The factors $\alpha$ and $1 - \alpha$ in the Fermi and Bose fluxes lead to effective rate constants $\alpha k_{T,j}$ and $(1-\alpha)k_{T,j}$, respectively"; 24) "Correspondingly, as $\alpha \to 1$, the Fermi and Bose fluxes separate (Fig. 5), so that the time scale for the Bose flux dramatically increases, which results in a sharp increase in the MFPT (Fig. 4)"; 25) "The obtained equation describes the time evolution of a system with mixed Bose-Fermi statistics which are controlled by the fraction of Fermi moves. The latter plays a role of an ET factor, i.e., at $\alpha = 1$ (Fermi moves) the system samples the state space, favoring the states that have not been visited yet"; and, finally, 26) "In particular, factors

$[1 \pm N_m(t)]$ in Eq. (7) (signs "+" and "−" are for the Bose and Fermi moves, respectively, and $m = i, j$) can be written as $\{N_{lim} \pm Q[N_m(t)]\}$, where $N_{lim}$ is the limiting population of the states for Fermi moves and $Q(N_m)$ is a function of $N_m$ [20]"
are incorrect in their parts concerning the Fermi statistics, Fermi-like moves, Fermi moves, Fermi effect and Fermi flux.

16. The equilibrium distribution given by Eqs. (15) (see Eq. (8) [1]) does not correspond to the mixed Bose-Fermi statistics.

17. The title "Mixed Bose-Fermi statistics: Kinetic equation and navigation through a network" does not correspond to the content of [1].

18. The incorrect kinetic equations (1), (6), (13) and (17) which were considered in [1] must be replaced by the following correct ones:

$\frac{dN_i}{dt} = \sum_{j, j \neq i} [k_{i,j} N_j(t) - k_{j,i} N_i(t)]$,

$\frac{dN_i}{dt} = \frac{1}{2} \sum_{j, j \neq i} L_i L_j \{k_{i,j}[1 - N_i(t)]N_j(t) - k_{j,i}[1 - N_j(t)]N_i(t)\}$

$+ \frac{1}{2} \sum_{j, j \neq i} \{k_{i,j}[1 + N_i(t)]N_j(t) - k_{j,i}[1 + N_j(t)]N_i(t)\}$, (23)

$\frac{dN_i}{dt} = \alpha \sum_{j, j \neq i} L_i L_j \{k_{i,j}[1 - N_i(t)]N_j(t) - k_{j,i}[1 - N_j(t)]N_i(t)\}$

$+ (1 - \alpha) \sum_{j, j \neq i} \{k_{i,j}[1 + N_i(t)]N_j(t) - k_{j,i}[1 + N_j(t)]N_i(t)\}$, (24)

$\frac{dN_i^{(m)}}{dt} = \alpha^{(m)} \sum_{j, j \neq i} \sum_{m'} L_i L_j \{k_{i,j}[1 - N_i^{(m)}(t)]N_j^{(m')}(t) - k_{j,i}[1 - N_j^{(m')}(t)]N_i^{(m)}(t)\}$

$+ (1 - \alpha^{(m)}) \sum_{j, j \neq i} \sum_{m'} \{k_{i,j}[1 + N_i^{(m)}(t)]N_j^{(m')}(t) - k_{j,i}[1 + N_j^{(m')}(t)]N_i^{(m)}(t)\}$, (25)

respectively.

Eqs. (23), (24) and (25) are the correct kinetic equations for the mixture of the Bose and Fermi moves.

19. We have from Eq. (24) that the mean numbers $N_{0i}$ according to the conditions of detailed balance obey the equation

$k_{i,j}[\alpha L_i L_j (1 - N_{0i}) N_{0j} + (1 - \alpha)(1 + N_{0i}) N_{0j}]$

$= k_{j,i}[\alpha L_i L_j (1 - N_{0j}) N_{0i} + (1 - \alpha)(1 + N_{0j}) N_{0i}]$. (26)

It is easy to establish from Eq. (26) that if $\alpha \geq 1/2$, then

$N_{0i} = \frac{1}{C_0 \exp(\beta E_i) + 2\alpha - 1}$, (27)

where a positive constant $C_0$, which depends on temperature and $N_{tot}$, is defined from the condition

$\sum_i N_{0i} = N_{tot}$. (28)

If $N_{0i} > 1$ for all states, then one can obtain from Eq. (26) that

$N_{0i} = \frac{1}{C_1 \exp(\beta E_i) - 1}$, (29)

where $C_1$ is defined from the condition (28). We conclude from Eq. (29) that the equilibrium distribution $N_{0i}$ does not depend on $\alpha$.

## Conclusion

It is shown that the kinetic equations considered in [1], equilibrium distribution obtained in [1], and results and conclusions obtained on the basis of the kinetic equation derived in [1] do not correspond to the mixed Bose-Fermi statistics, and, hence, the title "Mixed Bose-Fermi statistics: Kinetic equation and navigation through a network" does not correspond to the content of [1]. It is also shown that the kinetic equation corresponding to the case when the copies of the system are characterized by different values of the fraction of the Fermi-like moves (Eq. (9) [1]) is incorrect.

The correct kinetic equations (Eqs. (23), (24) and (25)) for the mixture of the Bose and Fermi moves are presented and the equilibrium distribution for the case when the probability of the Fermi moves is higher or equal to that of the Bose moves is obtained.